# The "Hydrogen Epoch of Reionization Array" (HERA) - Improvement of the antenna response with a matching network and scientific impacts

N. Fagnoni[1]   E. De Lera Acedo[1]

*Abstract* - **The Hydrogen Epoch of Reionization Array (HERA) is a new powerful radio-telescope, dedicated to the study of the early universe. Its main goal is to characterise the period of the universe where the first galaxies and stars started to form, by studying the evolution of the 21-cm emission signal from neutral hydrogen, during the "Epoch of Reionization". In this article, we present an electromagnetic and electrical co-simulation of the antenna performed with CST. We focus our analysis on the characterisation of the chromatic effects caused by the antenna, in particular multiple reflections of the received signal between the feed and the dish. These reflections can have an important impact on the scientific results, and it is crucial to keep them as low as possible. Therefore, we are currently developing a matching circuit which aims to improve the impedance matching between the feed and the front end.**

## 1 INTRODUCTION

Before the formation of the first galaxies, our universe was transparent and mainly composed of clouds of neutral hydrogen, which emitted photons with a wavelength of 21 cm (1420 MHz), when the spin configuration of the nucleus and electron flipped, passing from parallel to antiparallel. Therefore, during the "Dark Ages" of our universe, until about 300 million years after the Big Bang, the background emission was dominated by the 21-cm signal from this hyperfine transition. Then, the light of the first quasars and stars started ionizing the interstellar medium, and as a consequence, the 21-cm signal progressively disappeared. Thus, the study of the evolution of this signal during the "Epoch of Reionization" (EoR) will allow us to better understand the formation of the first structures of our universe. HERA is going to focus on the detection of this signal between 100 – 200 MHz, taking into account the redshift of the signal due to the expansion of the universe, which will allow us to probe a period between 400 million and 1 billion years after the Big Bang. The first elements of HERA are currently being built in the Karoo desert, in South Africa, and eventually this radio-interferometer will comprise more than 350 antennas, close-packed in a hexagonal grid [1].

The dish and feed are crucial and challenging elements, which must be designed very carefully, in order for HERA to achieve its ambitious science goals. Thus, they have been modelled with the simulation software CST, in order to characterise the antenna response [2]. The model used in our electromagnetic simulation is presented in section 2. In particular, the results highlight some chromatic effects caused by the antenna. We noticed some unwanted "ripples" on the received signal, due to the fact that the signal undergoes multiple reflections between the feed and the reflector. The main difficulty is that most of the EoR signal is buried in the foreground signal which is about 5 orders of magnitude more intense. However, as we will see, it is possible to recover a part of the EoR signal, on condition that its contamination by the foreground is limited, which implies to minimise the additional reflections. Ideally, the reflected signals should be attenuated by 60 dB minimum, 60 ns after the reception of the first signal [1], [3], [4].

Thus, in order to minimize these additional reflections, we are developing an electrical circuit, to improve the impedance matching between the antenna and the RF front-end. We will see in section 3 the effects of this matching circuit. Finally, in section 4, we will describe the impacts of these reflections on the characterisation of the power spectrum of the EoR signal, and we will see how this matching network can push the limits of the detection of this signal.

## 2 SIMULATION OF THE ANTENNA RESPONSE

### 2.1 Description of the HERA dish

The HERA dish can be divided into 3 major elements: the reflector, the dipole, and the cylindrical cage located just above the feed [5].

The reflector is a faceted paraboloid made up of 24 segments in aluminium, has a diameter of 14 m, with a focal ratio of 0.32, and is about 2.7-m high. At its vertex, a 1-m diameter hub made of concrete firmly holds the structure of the antenna. Based on the design of the radio-telescope PAPER, also in the Karoo desert, the feed is a simple crossed dipole antenna made of copper. Each dipole measures about 130 cm and is surrounded by two metallic discs, in order to broaden the frequency response of the antenna. The feed of HERA is surrounded by a cylindrical cage which protects it from external reflections coming from the close adjacent antennas. Finally, the cylindrical structure plus the feed are suspended by cables, 4.9 m above the dish [5]. The figure 1 shows the antenna modelled with CST.

---

[1] Astrophysics Group, Cavendish Laboratory, University of Cambridge, JJ Thomson Avenue, Cambridge CB3 0HE, United Kingdom
e-mail: nf323@mrao.cam.ac.uk, eloy@mrao.cam.ac.uk

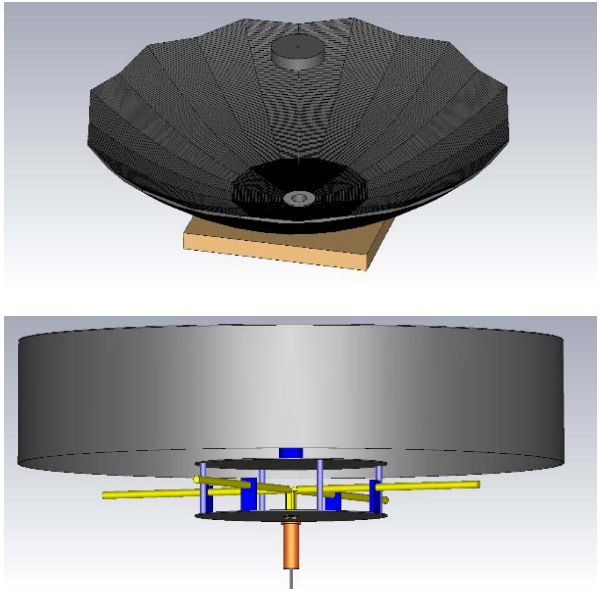

Figure 1. CST simulation of the HERA antenna.

**2.2 Electromagnetic and electrical co-simulation**

In order to have an accurate simulation, the effects of the RF front-end have to be taken into account. For the moment, HERA is still using the analogue chain developed for PAPER, which consists of an active balun with a pre-amplification stage, a transmission line and a receiver board.

Each arm of a dipole is terminated by a feeding bar directly connected to the electrical inputs of the balun, where the signal is amplified by a cascade of three amplifiers, and the antenna impedance transformed to 50 Ω with the help of a transformer. The overall gain of the balun is about 23 dB. Then a 180º hybrid coupler combines the two 180º out of phase signals from the dipole arms, and the output unbalanced signal is transmitted to the receiver via a standard RG6 coaxial cable. For the moment, 35-m cables are used, which causes an attenuation of about 3 dB. Finally, the receiver consists of a cascade of 5 amplifiers separated by 3-dB attenuators for inter-stage isolation, which gives an additional gain of about 75 dB. The impedance matching between the 75-Ω coaxial cable and the 50-Ω receiver is also performed with a transformer. In addition, the receiver includes a passband filter, with a bandwidth between 110 and 190 MHz, cantered on 150 MHz, and modelled by an elliptic filter for its very sharp roll-off. All these elements were carefully simulated with the software Genesys [6], thanks to the schematics of PAPER. The simulation shows that the global gain of the chain is about 95 dB.

Once the antenna has been modelled and the S-parameters of the RF front-end simulated, it is possible to run an electromagnetic and electrical co-simulation with CST. Indeed, the additional reflections which occur between the feed and the reflector are a direct consequence of the impedance mismatch which exists between the antenna and the front-end. Thus, a part of the signal is reflected back towards the feed, and so is radiated by the dipoles, before being reflected again by the dish. In our case, the analysis of the impedance of the antenna and the balun showed us an important disparity. In order to study the antenna response between 100 MHz and 200 MHz, a Gaussian pulse centred on 150 MHz with a bandwidth of 100 MHz, is used to excite a plane wave, coming from the zenith. The co-simulation is run using the transient solver of CST based on the "Finite Integration Technique" which discretizes and solves the Maxwell's equations in their integral form. The antenna response is simulated for 500 ns, with a time step of 0.003 ns.

**2.3 Antenna response**

Since the input and output signals are now available, it is possible to calculate the antenna response in the time and frequency domain. The output signal *out(t)* is equal to the input signal *in(t)* convolved by the antenna response *r(t)*. Then, thanks to the convolution theorem, this equation can be transformed in the frequency domain, and the antenna frequency response *R(f)* is simply the ratio of the Fourier transform of the output signal over the Fourier transform of the input signal:

$$R(f) = \frac{FT[out(t)]}{FT[in(t)]}$$

By taking the inverse Fourier transform, we can also obtain the antenna response in the time domain [2]. The figure 2 shows the simulated output signal. It is interesting to notice the additional ripples which do correspond to the round trip between the feed and the reflector. The figure 3 shows the spectrum of the input and output signals, as well as the antenna response. We can see that the additional reflections modify the spectral structure of the output signal.

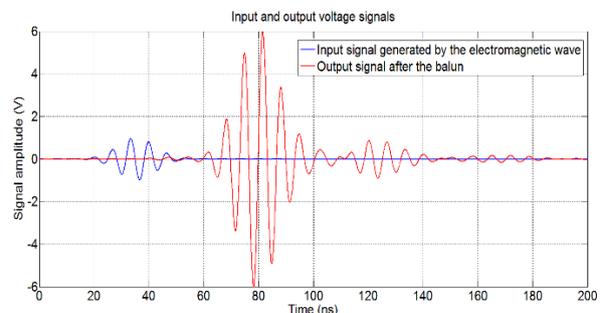

Figure 2. Input and output signals after the balun.

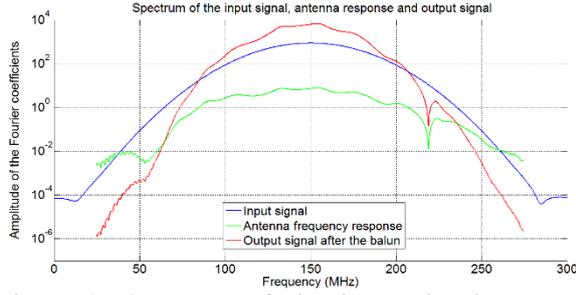

Figure 3. Spectrum of the input signal, antenna response and output signal, no matching network.

## 3 DEVELOPMENT OF AN IMPEDANCE MATCHING NETWORK

### 3.1 Electrical circuit

The impedance matching between the dipoles and the balun must be improved. However, impedance matching is a complex problem to solve, in particular when one needs to reach a high level of matching over a wide frequency band, considering impedances with resistive and reactive parts, and which significantly vary. Thus, this matching network was generated using the software Optenni Lab, with the antenna impedance and the S-parameters of the balun [7]. The circuit consists of a series of 10 lumped elements (capacitors and inductors), to be inserted between the feeding bars of the dipoles and the balun.

### 3.2 Improvement of the antenna response

The figure 4 shows a comparison between the normalized amplitude of the input signal, and output signals with and without matching network. To make the comparison easier, the signals were also shifted so that their maximum coincides at t=0. Thus, we can see that the overall level of reflection has been decreased by more than 10 dB. Concerning the frequency domain, the spectrum appears also slightly smoother. Finally, by looking at the S11 parameter of the balun on the figure 5, we can clearly see that the level of power reflected at the interface with the balun has been significantly decreased.

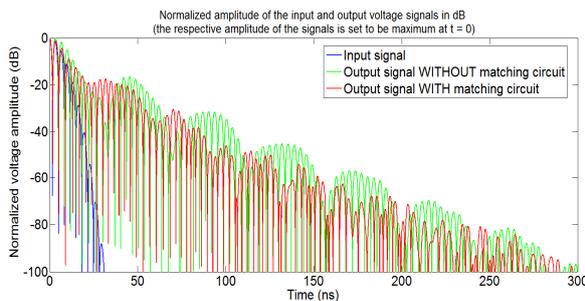

Figure 4. Normalized amplitude of the input and output voltages, with and without matching network.

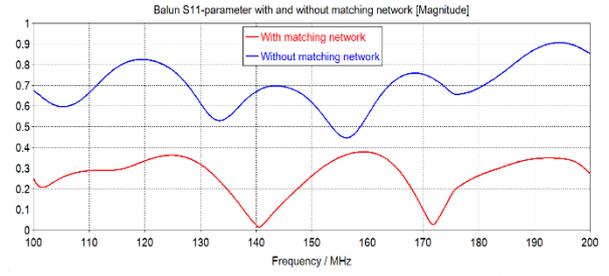

Figure 5. S11-parameter of the balun, with (blue) and without (green) matching network.

### 3.3 Attenuation, noise and sensitivity

However, adding another circuit has also some negative aspects. Each inductor and capacitor induces losses which were taken into account. The simulation of the matching circuit, and in particular its S21-parameter, shows that it causes an attenuation of the signal between 0.2 and 0.9 dB (see figure 6).

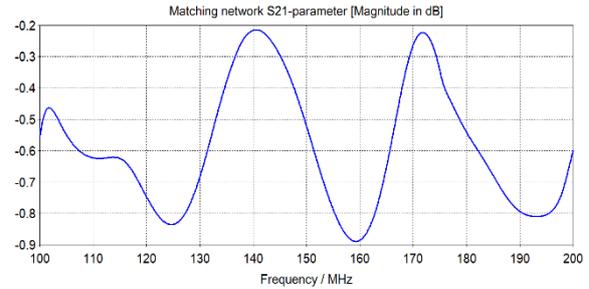

Figure 6. S21-parameters of the matching network.

Moreover, as a resistive and passive element, the circuit generates noise, which directly depends on its attenuation. But there is a second effect to consider: by changing the input impedance seen by the first amplifier, we also modify its noise figure $F_{amp}$:

$$F_{amp} = F_{min} + \frac{R_N}{G_S}|Y_s - Y_{opt}|^2$$

with $F_{min}$ the minimum noise figure of the amplifier, $R_N$ its equivalent noise resistance, $Y_{opt}$ the optimum source admittance which minimizes the noise figure, $Y_S$ the source admittance (so the matching network) and $G_S$ its real part. Consequently, a modification of the input impedance of the amplifier will result in a modification of the noise its generates. This effect may not be negligible, and so far we only tried to optimize the reflection of the input signal. Thus, our next work will be to try to estimate the impact of this modification on the sensitivity of the telescope, and if necessary find a trade-off between impedance matching for reflection and noise.

## 4    IMPACT ON THE SCIENCE DATA

HERA will observe the sky pointing to the zenith, in a "drift-scan" mode, with the goal to detect the EoR signal. During its observations, HERA will receive the signals from different patches of cold sky, which then will be "aggregated" in order to characterise its power spectrum. The distribution of the reionization is assumed to be rather isotropic across the sky at large scale, and so statistically, the features of the power spectrum should be the same for the considered scales. However, the main problem is that the EoR signal is "contaminated" by the foreground signal, from Galactic synchrotron emission and extragalactic radio sources. But, there are two different approaches to go around this problem: we can try either to "subtract" this foreground signal, but it requires a perfect knowledge of it, or try to observe a specific window of the EoR signal which is not affected by the foreground. As a first approach, this second option is preferable.

In radio-interferometry, the signals from two different antennas, forming a "baseline", are usually combined to get information about the sky intensity or "brightness". More precisely, an electromagnetic wave will be detected by the first antenna of the baseline, generating a voltage $v_i(t)$, and after a delay $\tau_s$ which depends on the length of the baseline and on the angle of arrival of the wave, the same signal will be detected by the second antenna, generating a voltage $v_j(t)=v_i(t-\tau_s)$. Then these two signals are cross-correlated. First, the time averaged product of the Fourier transforms of the voltage signals, $V_i(f)$ and $V_j(f)$, is calculated, giving the visibility $V_{ij}(f)$:

$$V_{ij}(f) = <V_i(f).V^*_j(f)>_t$$

Then the cross-correlation of the voltage signals, defining the "delay spectrum" $v_{ij}(\tau_s)$, can be obtained by taking the inverse Fourier transform of $V_{ij}(f)$, expressed as a function of the delay $\tau_s$:

$$v_{ij}(\tau_s) = \int V_{ij}(f).e^{2i\pi f \tau_s}.df$$

Finally, the "sky delay power spectrum", $P$, based on the delay spectrum and which contains the signal of the EoR affected by the foreground and the antenna chromaticity, can be approximated using the following formula [1]:

$$P \approx \left[\frac{\lambda^2}{2k_B}\right]^2 . \left[\frac{X^2Y}{\Omega B}\right] . v_{ij}(\tau_s)^2$$

with $\lambda$ the considered wavelength, $k_B$ the Boltzmann constant, $\Omega$ the integrated antenna beam, B the bandwidth, X and Y cosmological parameters. The most import here is to notice that the sky power spectrum can be expressed as a function of the delay.

In addition, the EoR and foreground spectrums are expected to be quite different: between 100 and 200 MHz, the foreground spectrum is rather smooth, unlike the EoR spectrum. As a consequence, we can show that the foreground power spectrum can be isolated in a specific delay window. Thus, for a perfect achromatic antenna, and for a given baseline, the maximum delay where the foreground spectrum is present should not exceed $\tau_{sMaxGeo} = b/c$, which is actually the maximum geometric signal delay associated to a baseline $b$, and also called "horizon delay limit". On the other hand, the weak EoR power spectrum is expected to spread beyond this limit. Thus, after $\tau_{sMaxGeo}$, the EoR power spectrum should be dominant [1], [4]. However, any chromatic effects caused by the antenna, like multiple reflections, will repeat the foreground signal, and so spread it in the delay domain, reducing the available information about the EoR spectrum. So it is crucial to minimise these reflections, hence the use of our matching network. In Thyagarajan, N., et al. (2016) [4], simulations of the foreground delay power spectrum obtained from achromatic and chromatic antenna models clearly illustrates the impact of the spillover of the foreground across the delay. Finally, by simulating a theoretical EoR delay power spectrum and comparing it with the level of its foreground model, Thyagarajan managed to determine what would be the constraints, in terms of attenuation of the reflected signals and reflection duration, to keep the foreground spillover below the EoR signal. By comparing his results to the figure 4, we found out that without matching network, the EoR delay power spectrum may be detected for delays $\tau_s$ above 300 ns, whereas with the matching network it would be possible to detect it after 200 ns, for a redshifted EoR signal at 150 MHz. To improve our results, this study should be repeated for different baseline lengths and frequencies, and by using the same type of foreground signal in our CST simulation. Nevertheless, this example shows the benefit of this tuned matching network, which concretely should allow us to get more spatial and spectral details about the structure of the EoR, by "unlocking" its power spectrum for lower delays.

## References

[1] DeBoer, D., et al. 2016, Hydrogen Epoch of Reionization Array (HERA), in preparation
[2] https://www.cst.com
[3] Ewall-Wice, A., et al. 2016, arXiv e-prints, arXiv: 1602.06277 [astro-ph.CO]
[4] Thyagarajan, N., et al. 2016, arXiv e-prints, arXiv: 1603.08958 [astro-ph.CO]
[5] DeBoer, D., 2015, HERA Phase I Feed Design
[6] http://www.keysight.com
[7] https://www.optenni.com